\documentclass{aa} 

\usepackage{graphicx}
\usepackage{txfonts}
\begin{document}
\title{Simultaneous polarization  monitoring         of     supernovae
  \object{SN\,2008D/XT\,080109}   and  \object{SN\,2007uy}: isolating
  geometry from dust}


\author{J.  Gorosabel\inst{1} \and A. de Ugarte Postigo\inst{2,3} \and A.~J. Castro-Tirado\inst{1} \and I. Agudo\inst{1,4} \and M. Jel\'{\i}nek\inst{1} \and S. Leon\inst{5} \and T. Augusteijn\inst{6} \and J.~P.~U.~Fynbo\inst{7} \and J. Hjorth\inst{7} \and  M.~J. Micha{\l}owski\inst{7} \and  D. Xu\inst{7} \and P. Ferrero\inst{8,9,10} \and D.~A. Kann\inst{10} \and S. Klose\inst{10} \and A. Rossi\inst{10} \and J.~P. Madrid\inst{11} \and A.~LLorente\inst{12} \and M. Bremer\inst{13} \and J.-M. Winters\inst{13}}

\institute{Instituto de Astrof\'{\i}sica de Andaluc\'{\i}a (CSIC),  Glorieta de la Astronom\'{\i}a s/n, 18008 Granada, Spain. \and  European Southern Observatory, Casilla 19001, Santiago 19, Chile. \and INAF/Osservatorio Astronomico di Brera, via Bianchi 46, 23807 Merate, LC, Italy. \and Institute for Astrophysical Research, Boston University, 725 Commonwealth Avenue, Boston, MA 02215, USA. \and Joint ALMA Observatory, Av. El Golf 40, Piso 18, Las Condes, Santiago de Chile, Chile. \and Nordic Optical Telescope, Apartado 474, 38700 Santa Cruz de La Palma, Spain. \and Dark Cosmology Centre, Niels Bohr Institute, University of Copenhagen, Juliane Maries Vej 30, 2100 Copenhagen {\O}, Denmark. \and  Instituto de   Astrof\'{\i}sica de  Canarias,  V\'{\i}a L\'actea  s/n, 38200  La Laguna,  Tenerife,  Spain \and Departamento de Astrof\'{\i}sica, Universidad de La Laguna (ULL), E-38205 La Laguna, Tenerife, Spain. \and Th\"uringer Landessternwarte Tautenburg, Sternwarte 5, D-07778 Tautenburg, Germany. \and Centre for Astrophysics \& Supercomputing, Swinburne University of Technology, P.O. Box 218, Hawthorn, VIC 3122, Australia. \and INSA, Herschel   Science  Operations  Centre,  European  Space Agency,   Villafranca del  Castillo,  PO   Box   50727, 28080  Madrid, Spain. \and  Institute de Radioastronomie Millim\'etrique  (IRAM), 300 rue de la Piscine, 38406 Saint Martin d\'\rm H\'eres, France.}

\date{Received May 14, 2010}

\abstract  {{The possible  existence  of a  continuum encompassing the
    diversity of   explosive  stellar deaths,  ranging   from ordinary
    supernovae  (SNe; lacking any sign  of  a relativistic outflow) to
    relativistic   hypernovae associated with  energetic long duration
    gamma-ray   bursts  (GRBs), is under    intense  debate.  In  this
    context, the  supernova   \object{SN\,2008D} associated with   the
    X-ray  transient    \object{(XT)   080109}   could    represent  a
    paradigmatic case, since it might exemplify a potential borderline
    transition event.   Optical  polarimetric studies could contribute
    to shed  light  on the  different  interpretations given  in   the
    literature      for this        supernova              (hereafter,
    \object{SN\,2008D/XT\,080109}).}}{The  main  aim    is    to infer
  geometric information of  SN\,2008D/XT\,080109 through  the study of
  the evolution of  its  linear  optical  polarization.  \rm  We  also
  report  the  polarization   evolution of  \object{SN\,2007uy},   and
  discuss the properties of  the host galaxy interstellar medium (ISM)
  towards the  XT.   The final   goal is  to compare the  polarization
  properties, and therefore the geometries, of both SNe.}{We present a
  $V$-band linear   polarization monitoring  campaign  carried out for
  \object{SN\,2008D/XT\,080109} and \object{SN\,2007uy},   which shone
  for weeks  contemporaneously in \object{NGC\,2770}.   This fortunate
  coincidence  brought  us the   opportunity  to observe  both objects
  simultaneously,  and most  importantly,  with identical instrumental
  setups.  The observations  span 74.9 days,  starting 3.6  days after
  the  XT and are distributed in  11 visits.  In addition we performed
  observations  in the millimetre (mm)  range in order to identify the
  dominant origin of the  observed polarization.}  {We report positive
  linear    polarization   detections     at   several   epochs    for
  \object{SN\,2008D/XT\,080109}     at  a level     of $\sim$1\%.  For
  \object{SN\,2007uy} the measured polarization is around $\sim$1.5\%.
  In  both cases the  observed linear  polarization seems dominated by
  the host galaxy interstellar polarization (HGIP), especially for the
  case  of  \object{SN\,2007uy}.     \object{SN\,2007uy} shows  Stokes
  parametres   consistent  with no  time    evolution, which  could be
  described by the HGIP plus a constant  eccentricity expansion on the
  sky   plane.   Over    the    course   of   our   observations    of
  \object{SN\,2007uy}, we find that its total polarization signal does
  not   change by more  than  0.29\% with  a 90\% confidence interval.
  Despite   the  dominant   HGIP,  a  statistical   analysis   of  the
  distribution of the \object{SN\,2008D/XT\,080109} Stokes  parametres
  suggests   that it    could   show a  possible   intrinsic  variable
  polarization component.  Moreover,  assuming the polarization signal
  from \object{SN\,2007uy} is   constant,  we find  that the  temporal
  evolution     of   the     intrinsic   \object{SN\,2008D/XT\,080109}
  polarization could  be  explained   by  an   aspherical axisymmetric
  expansion  with  variable eccentricity,  although other more complex
  geometric scenarios are also compatible  with the data.  We come  to
  the  same  result    even  if  we   make   no  assumption   on   the
  \object{SN\,2007uy}   Stokes   parametres,  although   at    a lower
  significance level. }  { We conclude that the data seem to suggest a
  potential  symmetry axis for \object{SN\,2008D/XT\,080109}, which is
  reinforced  when \object{SN\,2007uy}  is  assumed  to  have constant
  Stokes  parametres and used as  reference star.   We suggest that at
  least  the  projected,     if   not  the   intrinsic,   geometry  of
  \object{SN\,2008D/XT\,080109}  and     \object{SN\,2007uy}  could be
  different. } {}


 \authorrunning{Gorosabel et al.}
 \titlerunning{Simultaneous polarization of \object{SN\,2008D/XT\,080109} and  \object{SN\,2007uy}}

   \maketitle

%

\section{Introduction}

The    X-ray   transient \object{(XT)   080109}  was   serendipitously
discovered  on 2008 January 9.56446 UT  by  \textit{Swift} as a bright
X-ray outburst during  a scheduled observation  of \object{SN\,2007uy}
(Berger \& Soderberg 2008;  Kong \& Maccarone 2008).  Optical follow-up
observations revealed the presence of a supernova (SN) associated with
the       XT  (Deng      \&        Zhu    2008,  hereafter       named
\object{SN\,2008D/XT\,080109}).          \object{SN\,2008D/XT\,080109}
underwent  a transition  1-2  weeks after the XT  from  Type Ic  to Ib
(Malesani et al.  2009;  Blondin et al.  2008;  Valenti et al.   2008;
Modjaz et  al.  2009),  resembling the   spectral evolution   seen  in
\object{SN\,2005bf} (Anupama et al. 2005; Modjaz et al. 2005; Tominaga
et  al.    2005;     Wang      \&    Baade     2005).      Eventually,
\object{SN\,2008D/XT\,080109}   was  spectroscopically classified as a
He-rich Ib supernova (Modjaz et al. 2009).

The origin of this XT is still unclear.  Two  main scenarios have been
proposed to explain its nature; the  stellar shock breakout (Soderberg
et al. 2008; Chevalier \& Fransson 2008) and the X-ray flash framework
(Li 2008; Xu et al. 2008; Mazzali et al. 2008).  In the first case the
early emission would be entirely  explained by the supernova,  whereas
in the second case a relativistic  outflow from a central engine would
be  required.  Radio  data seem  to  indicate no superluminal motions,
arguing   against the presence  of   a long-lived relativistic outflow
(Bietenhloz et al.  2009), therefore  supporting the first  framework.
However, different spectral analyses of  the X-ray data did not  yield
conclusive  results  on the possible  presence  of a thermal component
expected from the shock breakout  wave (Chevalier \& Fransson 2008; Li
2008; Soderberg et al. 2008; Modjaz  et al.  2009).  Within the second
scenario  \object{SN\,2008D/XT\,080109} was  proposed as  a transition
between the most    energetic  hypernovae (linked  to  canonical  long
duration gamma-ray bursts) and  standard core-collapse SNe (Mazzali et
al.        2008).      In  fact       it      was   suggested     that
\object{SN\,2008D/XT\,080109} is a side-viewed, bipolar explosion with
a viewing angle of $>50^\circ$ from the polar direction (Tanaka et al.
2009a).   It should  be   noted  that spectroscopic  observations   of
\object{SN\,2008D/XT\,080109}   suggested  an   aspherical   expansion
geometry (Modjaz et al. 2009; Tanaka et al. 2009a).

\object{SN\,2007uy} was discovered on 2007 Dec.   31.669 UT (Nakano et
al. 2008). It was classified as an ordinary Ib SN (with no hint of any
superluminal  outflow) before   the light  curve  maximum  (Blondin \&
Calkins 2008), showing  spectral similarities with \object{SN\,2004gq}
(Pugh   et al. 2004;   Modjaz  \& Falco  2005).   In  the optical both
\object{SN\,2008D/XT\,080109}      and     \object{SN\,2007uy}   shone
contemporaneously  for weeks in   the nearby ($z=0.007$) spiral galaxy
\object{NGC\,2770}.  This   fortunate   coincidence  brought   us  the
opportunity  to  observe  both    objects   simultaneously,  and  most
importantly, with the same instrumental setup.

Recently   Maund et al.    (2009)  have reported a spectropolarimetric
study of \object{SN\,2008D/XT\,080109}  based on data acquired on 2008
January  31.22 and February  15.18 UT.  One of  the main advantages of
spectropolarimetry of SNe  with  respect to broad-band  polarimetry is
its  ability to  infer  geometric  and dynamical  information for  the
different  chemical constituents    of  the   explosion.    Broad-band
polarimetric observations construct  a rougher picture of  the stellar
death,   but  require      lower     signal-to-noise   ratios     than
spectropolarimetry.  Hence broad-band polarimetric observations can be
extended to objects at higher  redshifts or/and they allow to  enhance
the polarimetric coverage and  sampling of the light curve, especially
at epochs far from the maximum when the SN is dimmer.

The dominant intrinsic polarization in  SNe is thought to originate by
Thompson   photon scattering    through   an  aspherical  photospheric
expansion (H\"oflich  1991).  If the SN  photosphere projection on the
sky    plane is   an  ellipse, in     general a  non-cancelled  linear
polarization is  expected perpendicular or parallel  to the major axis
(Kasen et al. 2003).  Therefore, if  the expansion ellipticity evolves
with time, keeping the direction   of  both axes unchanged, then   the
Stokes parametres should  also move on a straight  line of the  Stokes
plane.  Once  the Milky Way  Galactic  interstellar polarization (GIP)
has been  corrected, this straight line  is  expected to  be displaced
from the  origin   of  the Stokes   plane   due to   the host   galaxy
interstellar polarization  (HGIP).   Using  several  different methods
Maund et al.  (2009) estimated a HGIP of $P\sim1.2$\% and $\theta \sim
134^{\circ}$   for \object{SN\,2008D/XT\,080109}.  The contribution of
additional   polarization  components (variable   line   polarization,
irruption of the stellar core, variable  HGIP, etc.)  can distort this
rough   ellipsoidal  stellar picture,   resulting in   a more  complex
geometrical configuration   (see  Maund et al.   2009,  and references
therein).  Asphericities  in explosive stellar  deaths tend to be low,
usually yielding optical polarization values  below $\sim1$\% (Wang \&
Wheeler 2008).  However, counterexamples  exist, such  as the Type  Ic
\object{SN\,1997X}      (Wang         et    al.        2001)        or
\object{XRF\,060218/SN\,2006aj} (Gorosabel  et al. 2006;  Maund et al.
2007).

In   this  paper we   report  optical   polarimetric observations   of
\object{SN\,2008D/XT\,080109}     and  \object{SN\,2007uy} aimed    at
obtaining geometrical  information  on the expansion of  these events.
In addition we show the results of  millimetre (mm) observations which
were used  to infer information about the  host  galaxy extinction and
consequently about the HGIP impact on the observed polarization.


\section{Observations and data analysis}

\begin{table*}
\begin{center}
\caption{\label{table1} Polarimetric observations of \object{SN\,2008D/XT\,080109} and \object{SN\,2007uy}.}
\begin{tabular}{cccc|ccc|ccc}
\hline
\hline
Date$^a$ &\hspace{-0.5cm}   Telescope   &\hspace{-0.3cm} Exposure  &\hspace{-0.4cm}           Seeing &                            \multicolumn{3}{c}{\object{SN\,2008D/XT\,080109}}&                      \multicolumn{3}{|c}{\object{SN\,2007uy}$^b$}\\
2008, UT &\hspace{-0.5cm} (+Instrument) &\hspace{-0.3cm} time (s)  &\hspace{-0.4cm} ($\prime\prime$) & $P \pm \sigma_P $ (\%) &\hspace{-0.1cm} $\theta \pm \sigma_\theta$ (deg) & \hspace{-0.3cm}$V\pm\sigma_V^c$ & $P \pm \sigma_P$ (\%) &\hspace{-0.1cm} $\theta \pm \sigma_\theta$ (deg) &\hspace{-0.3cm} $V\pm\sigma_V^c$\\
\hline
\hline
Jan 13.1644 &\hspace{-0.2cm}NOT(+ALFOSC)&\hspace{-0.3cm} 12$\times$600 &\hspace{-0.4cm} 0.7 &  $0.95\pm0.20$ &\hspace{-0.1cm} $114.9\pm5.9 $ &\hspace{-0.3cm} $18.44\pm0.05$ & -------- & --------  &  --------\\
\hline
Jan 13.2642 &\hspace{-0.2cm}VLT(+FORS1) &\hspace{-0.3cm} 2330$^d$ &\hspace{-0.4cm} 1.0 &  $0.68\pm0.22$     &\hspace{-0.1cm} $119.2\pm9.5  $&\hspace{-0.3cm} $18.39\pm0.05$ & $1.45\pm0.25$&\hspace{-0.1cm} $178.6\pm4.6$&\hspace{-0.3cm} $15.78\pm0.05$\\
            &            &          &     & ($0.65\pm0.33$)$^e$&\hspace{-0.1cm}($124.5\pm13.3 $)&               &($1.59\pm0.04$)&\hspace{-0.1cm}($177.4\pm0.7$)&             \\
\hline
Jan 15.0780 &\hspace{-0.2cm}NOT(+ALFOSC)&\hspace{-0.3cm} 12$\times$600 &\hspace{-0.4cm} 0.9 & $0.85\pm0.28$  &\hspace{-0.1cm} $106.1\pm9.4 $ &\hspace{-0.3cm} $18.28\pm0.05$ & -------- & --------   &  --------\\
\hline
Jan 17.2518 &\hspace{-0.2cm}VLT(+FORS1) &\hspace{-0.3cm} 8$\times$285  &\hspace{-0.4cm} 1.0 & $0.84\pm0.45$  &\hspace{-0.1cm} $138.3\pm8.1 $ &\hspace{-0.3cm} $17.90\pm0.05 $ & $1.43\pm0.14$&\hspace{-0.1cm} $182.5\pm5.1$  &\hspace{-0.3cm}  $15.69 \pm 0.05$\\
            &            &          &    & ($1.14\pm0.51$)&\hspace{-0.1cm}($141.0\pm7.2$) &                &($1.59\pm0.04$)&\hspace{-0.1cm}($177.4\pm0.7$)&             \\ 
\hline
Jan 30.1858 &\hspace{-0.2cm}VLT(+FORS1) &\hspace{-0.3cm} 4$\times$285  &\hspace{-0.4cm} 0.8 & $1.05\pm0.06$  &\hspace{-0.1cm} $135.3\pm1.7 $ &\hspace{-0.3cm} $17.31\pm0.05 $ & $1.64\pm0.06$&\hspace{-0.1cm} $178.5\pm1.1$  &\hspace{-0.3cm}  $16.39 \pm 0.05$\\
            &            &          &    & ($1.11\pm0.08$)&\hspace{-0.1cm}($134.0\pm2.0$) &                &($1.59\pm0.04$)&\hspace{-0.1cm}($177.4\pm0.7$)&             \\ 
\hline
Jan 30.2022 &\hspace{-0.2cm}VLT(+FORS1) &\hspace{-0.3cm} 4$\times$285  &\hspace{-0.4cm} 0.8 & $1.28\pm0.06$  &\hspace{-0.1cm} $132.5\pm1.4 $ &\hspace{-0.3cm} $17.27 \pm0.05 $ & $1.56\pm0.06$&\hspace{-0.1cm} $176.1\pm1.1$ &\hspace{-0.3cm}  $16.36\pm 0.05$\\
            &            &               &     & ($1.21\pm0.08$)&\hspace{-0.1cm}($133.3\pm1.9$) &                &($1.59\pm0.04$)&\hspace{-0.1cm}($177.4\pm0.7$)&             \\ 
\hline
Feb 11.9326 &\hspace{-0.2cm}CAHA(+CAFOS)&\hspace{-0.3cm} 13$\times$400&\hspace{-0.4cm} 2.0 & $1.58\pm0.97$  &\hspace{-0.1cm} $126.6\pm16.4 $&\hspace{-0.3cm} $17.82\pm0.09$ & $0.92\pm0.88$&\hspace{-0.1cm} $179.8\pm26.0$ &\hspace{-0.3cm} $17.09\pm0.11$\\
            &            &              &     & ($1.19\pm1.51$)&\hspace{-0.1cm}($131.6\pm20.5 $)&                &($1.59\pm0.04$)&\hspace{-0.1cm}($177.4\pm0.7$)&               \\ 
\hline
Feb 26.1107 &\hspace{-0.2cm}CAHA(+CAFOS)&\hspace{-0.3cm} 16$\times$600&\hspace{-0.4cm} 1.9 & $< 2.76^f$ & -------- &\hspace{-0.3cm}$18.52\pm0.09$& $1.81\pm0.63$& \hspace{-0.1cm}$173.5\pm14.4$ &\hspace{-0.3cm} $17.56\pm0.08$\\
            &            &              &     &($< 2.94^f$)& -------- &              &($1.59\pm0.04$)&\hspace{-0.1cm}($177.4\pm0.7$)&               \\
\hline
Feb 26.9444 &\hspace{-0.2cm}NOT(+ALFOSC)&\hspace{-0.3cm} 16$\times$400&\hspace{-0.4cm} 1.3 & $0.98\pm0.80$ &\hspace{-0.1cm} $124.6\pm23.2 $ &\hspace{-0.3cm} $18.79\pm0.08$ & --------     & --------       &  -------- \\
\hline
Mar 2.1102  &\hspace{-0.2cm}VLT(+FORS1) &\hspace{-0.3cm} 4$\times$525 &\hspace{-0.4cm} 1.1 & $1.42\pm0.46$ &\hspace{-0.1cm} $139.0\pm9.1 $ &\hspace{-0.3cm} $18.80\pm0.05$ & $1.83\pm0.19$ &\hspace{-0.1cm}$174.6\pm3.2$  &\hspace{-0.3cm} $17.55\pm0.05$\\
            &            &              &     &($1.17\pm0.51$)&\hspace{-0.1cm}($134.7\pm11.4$)&            &($1.59\pm0.04$)&\hspace{-0.1cm}($177.4\pm0.7$)&             \\
\hline
Mar 28.0207 &\hspace{-0.2cm}VLT(+FORS1) &\hspace{-0.3cm} 4$\times$525 &\hspace{-0.4cm} 1.1 & $0.75\pm0.57$ &\hspace{-0.1cm}$111.2\pm21.9$ &\hspace{-0.3cm} $19.42\pm0.05$ & $1.58\pm0.42$ &\hspace{-0.1cm}$179.6\pm7.6$ &\hspace{-0.3cm} $17.98\pm0.05$\\
            &            &          &     &($0.73\pm0.78$)&\hspace{-0.1cm}($112.7\pm21.0$)&                &($1.59\pm0.04$)&\hspace{-0.1cm}($177.4\pm0.7$)&             \\
\hline
\multicolumn{10}{l}{\scriptsize{$a$~~Mean observing epoch.}}\\

\multicolumn{10}{l}{\scriptsize{$b$~~For the NOT epochs \object{SN\,2007uy} was out of the
       FoV,   so the  degree  of   linear polarization  ($P$), the
       position angle ($\theta$), and the magnitude ($V$) could not be determined.}}\\

  \multicolumn{10}{l}{\scriptsize{$c$~~Calibration based on Malesani et al. (2009). The photometric   errors   include the  zero
      point uncertainty.}}\\

 \multicolumn{10}{l}{\scriptsize{$d$~~The   polarimetric   cycle   was
     interrupted and restarted several times.  The \object{SN\,2007uy}
     observing epoch and the exposure time are slightly different from
     the \object{SN\,2008D/XT\,080109} ones,}}\\

 \multicolumn{10}{l}{\scriptsize{~~~~~~ corresponding to Jan 13.2708 UT and 990 seconds, respectively.}}\\

  \multicolumn{10}{l}{\scriptsize{$e$~~The
  \object{SN\,2008D/XT\,080109} $P$ and $\theta$ values given for VLT
  and  CAHA    in     between   parentheses     assume    a   constant
  \object{SN\,2007uy}  polarization    given     by  $\langle        P
  \rangle=1.59\pm0.04$\% and $\langle \theta \rangle =177.4\pm0.7^{\circ}$.}}\\

  \multicolumn{10}{l}{\scriptsize{$f$~~1$\sigma$ upper limit.}}\\

\hline
\end{tabular}
\end{center}
\end{table*}

\subsection{Optical observations}
Table~\ref{table1} displays the log  of our optical observations.  All
the data  were taken in  the $V$-band.  The observations  were carried
out with the 2.2m Calar Alto telescope (CAHA), the 2.5m Nordic Optical
Telescope  (NOT) and  the  8.2m Unit  Telescope  2 of  the Very  Large
Telescope (VLT).

The NOT observations were performed with ALFOSC, through a calcite and
a 1/2 wave retarder  plate.  Four images  of the field were  acquired,
rotating    the  retarder   plate   at   $0^{\circ}$,  22.5$^{\circ}$,
45.0$^{\circ}$ and  67.5$^{\circ}$.   The  calcite  plates  reduce the
ALFOSC field of view (FoV) to 140$^{\prime \prime}$ in diameter (pixel
scale  of 0$\farcs$19/pix).  \object{SN\,2007uy}  was left  out of the
NOT images due to this reduced FoV.

The CAHA observations  were based on the CAFOS  instrument.  The CAFOS
polarization unit  uses a Wollaston  prism instead of a  calcite plate
and  has  a  strip  mask  on  the  focal  plane  to  avoid  accidental
overlapping on  the CCD.   The total  FoV of CAFOS  is composed  by 14
strips of $9^{\prime}\times18^{\prime\prime}$  each, providing a large
enough   FoV   to   image  both \object{SN\,2008D/XT\,080109}   and
\object{SN\,2007uy} simultaneously.

The  VLT observations were done  with FORS1,  with  a setup similar to
CAFOS.    As  in   CAFOS,  the  large  FoV   of FORS1   allowed  us to
simultaneously      image      \object{SN\,2008D/XT\,080109}       and
\object{SN\,2007uy}.   In  some of  the    FORS1 visits we  used   the
1$\times$1   bin  read-out  mode  in   order  to  avoid saturation  of
\object{SN\,2007uy}  while keeping a high  signal  to noise ratio  for
\object{SN\,2008D/XT\,080109}.  We note that,  in addition to  the CCD
binning, the  seeing and  transparency conditions varied significantly
from night   to  night.  This   made it impossible   to have  a bright
unsaturated field   star   common to  all   our   images.  Apart  from
\object{SN\,2008D/XT\,080109}, \object{SN\,2007uy} was the object most
frequently imaged, being present in the FoV of all the FORS1 and CAFOS
images.   This   fact provided us   with   the  opportunity  of  using
\object{SN\,2007uy}  as a  polarimetric reference  star  on the Stokes
plane (see discussion of Sect.~\ref{polaanalysis}).

The    images were  reduced  using standard   procedures running under
IRAF\footnote{IRAF is   distributed by the  National Optical Astronomy
  Observatory, which is  operated  by the Association  of Universities
  for  Research in Astronomy  (AURA)  under cooperative agreement with
  the National Science Foundation.}.   The master flat field image was
created   combining sky flat  field  frames taken without polarization
units in the light path.  This  process was applied separately for the
3 instruments.  In principle, the host  galaxy background might pose a
potential problem for the  photometric accuracy of \object{SN\,2007uy}
and      \object{SN\,2008D/XT\,080109}.        Inspection    of    the
\object{NGC\,2770}     region     around   each      SN    shows  that
\object{SN\,2007uy} is surrounded  by a more  inhomogeneous background
than \object{SN\,2008D/XT\,080109}.  Thus we double-checked the impact
of the   host  galaxy background   level  determination by  performing
aperture photometry with radii ranging from  0.5 to 2.5 times the FWHM
(Full   Width at Half Maximum),  and  varying the  inner  radii of the
background annuli from 3 to 5 times the FWHM.   The annuli widths were
also varied  from 2.5  to  4 times  the  FWHM.   We verified that  the
resulting  Stokes    parametres were consistent   within  error  bars,
independent of the apertures and  annuli used.  Then the apertures and
annuli yielding minimum  errors were adopted  (typically the apertures
and  the inner  radii  used  were the  FWHM  and  3-4 times  the FWHM,
respectively).   In any case,   in  order to account  for  potentially
dismissed photometric    error sources, the   statistical  analysis of
Sect.~\ref{variability} was duplicated considering also    photometric
errors augmented by 20\%.

For the ALFOSC    and CAFOS data the    determination  of the   Stokes
parametres was  done by   fitting  the $S(\theta)$ function   with the
corresponding Milky Way  GIP normalization factor (di Serego Alighieri
1997).    The GIP normalization  factor  was calculated using Galactic
field stars.

In the case of FORS1 the Stokes parametres were calculated via Fourier
expansion   (see expressions   4.5      and  4.6    of    FORS1
  manual\footnote{Doc.  N.   VLT-MAN-ESO-13100-1543.  Issue 82.1, date
    27/02/2008.}).   Then the Stokes parametres  of all the objects in
  the  images were    corrected for the   FORS1  spurious polarization
  correction, which considers the instrumental polarization dependence
  across              the            FoV                \footnote{{\tt
      http://www.eso.org/paranal/instruments/fors1/inst/pola.html}}.
  Next,  the $V$-band angle offset of  the 1/2 wave retarder plate was
  included (given in the Table 4.7  of FORS1 manual).  All the objects
  (including both SNe) of all FORS1 visits used the same angle offset,
  minimizing a potential source of  internal scattering in the  Stokes
  plane.   Finally, the GIP correction  was introduced by shifting the
  origin  of the  Stokes  plane  to    the  barycentre of  the   field
  stars\footnote{The FORS1 reduction  steps explained above follow
    the recommendations given by Dr. T.  Szeifert, priv. comm.}.

  Due to the reduced   ALFOSC FoV only   two unsaturated bright  field
  stars were  available   for the GIP    correction of the  NOT  data.
  Fortunately for the three NOT epochs the stars  used for the NOT GIP
  correction remained the same.  This  assures internal consistency of
  the position  of the  NOT   Stokes parametres  on the Stokes  plane.
  Thus, a  potential     relative  shift (owing    to different    GIP
  corrections) of the NOT Stokes parametres with respect to each other
  was  minimized.   This  fact will  be  of relevance   in the further
  discussion  since,    as will  be   argued  in Sect.~\ref{axis}, the
  polarimetric data  might suggest the  existence  of a symmetry  axis
  which is consistently determined by different instrumental setups.

The wider FoV of FORS1 and CAFOS provided a larger number of stars for
the GIP correction  than   with ALFOSC.  Due  to  different  observing
conditions (seeing, transparency and CCD binning) at  VLT and CAHA the
stars varied from one night to another.  This was not critical for the
FORS1   and CAFOS   images,   since   the  GIP  correction    for both
\object{SN\,2008D/XT\,080109}  and  \object{SN\,2007uy}   were  always
calculated using  the  same set of  stars  for each  epoch.  And  most
importantly, given that \object{SN\,2007uy}  was very well detected in
all the VLT and CAHA images, we could use this  object (justified by a
statistical analysis,   see Sect.~\ref{variability})   as  a secondary
calibrator and  keep     it  fixed   on   the   Stokes   plane    (see
Sect.~\ref{axis}).

It is  important to  stress that  the  GIP corrections applied  to the
different  data-sets were  very similar since  the Galactic extinction
towards \object{NGC\,2770}  is only $A_{V}=0.1$   mag (Schlegel et al.
1998; Cardelli et al.   1989).   So in  all cases the  GIP corrections
applied were low ($\sim0.2-0.3$\%),   in agreement with   the Galactic
polarization   predictions ($P_{max}  = 2.9\%  \times  A_V =  0.29$\%;
(Serkowski   et al. 1975; Cardelli   et al. 1989).    In fact, the GIP
corrections  were    always well below    the  typical  level   of our
polarization detections ($\sim 1-1.5 $\%).

Verification of the photometry, calibration and the reduction was done
by observation of polarimetric standards.  The standard stars observed
were \object{NGC\,2024-1} and \object{Vela1-95} at VLT (Fossati et al.
2007;  Whittet et al.   1992), \object{HD19820} at  CAHA (Wolff et al.
1996),   and \object{HD94851}, \object{BD25727}, \object{HD251204} and
\object{G191B2B} at NOT (Turnshek  et al. 1990;  Schmidt et al. 1992).
Table~\ref{table1}  shows the inferred  $\theta$  and $P$ values, once
they were corrected for the  GIP and the statistical bias (multiplying
$P$  by $\sqrt{1-(\sigma_P/P)^{2}}$\,) due to  the fact that  $P$ is a
positive   quantity      (Wardle   \&     Kronberg       1974).    The
\object{SN\,2008D/XT\,080109} $P$  and   $\theta$ values displayed  in
Table~\ref{table1}  for VLT and CAHA in  between  parentheses assume a
constant \object{SN\,2007uy}   polarization    given  by  $\langle   P
\rangle=1.59\pm0.04$\%     and          $\langle     \theta    \rangle
=177.4\pm0.7^{\circ}$.

For  some observing epochs  we explored maximising the time resolution
of our   polarimetric  monitoring at  the   expense  of  enlarging the
polarimetric errors.   This was done,  when possible, by splitting the
polarimetric  data of one  night in  cycles  of four images, i.e., the
minimum block  of   images  necessary to  get a   polarimetric  point.
Unfortunately  most of the    polarization detections  showed   modest
significance  levels (between $1.1  \sigma$ and  4.7  $\sigma$), so we
could not split the data.   Only the VLT data-set  of January 30 could
be divided in  two cycles of four   images, while still  keeping
high-significance polarization detections  (above  17.5  $\sigma$  for
each of the two cycles).  We verified that the  joint data acquired on
January  30  (composed of  two cycles  of  four  images) yields Stokes
parametres consistent  with the ones  obtained when  the data of  that
night  are split  in two halves.    This separation provided  an extra
high-quality data-point for  the  statistical analyses carried  out in
Sect.~\ref{variability},      Sect.~\ref{axis}              and
  Sect.~\ref{robust}.

The $V$-band polarimetry values synthesized by Maund et al. (2009) for
\object{SN\,2008D/XT\,080109} on January 31.22 and February 15.18 UT
bring us the opportunity to  cross-check our results.  Our two closest
observations, carried out on January  30.2022 and February 11.9326 UT,
yield  $P$ and $\theta$  values  fully  consistent with  Maund et  al.
(2009).     We are aware   that   this comparative exercise,  although
satisfactory, is  only limited to two  epochs close  to the  light curve
maximum     and exclusively focused on \object{SN\,2008D/XT\,080109},
since Maund et al.  (2009) did not observe \object{SN\,2007uy}.

\subsection{Millimetre observations}

\object{SN\,2008D/XT\,080109} was observed at  1.2\,mm on  January 25,
28, and 30, 2008 with the MAMBO~II bolometer on the IRAM 30m Telescope
(see   Table~\ref{table2}).   The   data were  reduced   following the
standard procedure with the {\it mopsic}  data reduction software.  In
order to  estimate  the \object{NGC\,2770} contribution  at 1.2\,mm to
our  previous  \object{SN\,2008D/XT\,080109} observations we measured,
on January 31, the flux of two adjacent regions (Adj1 and Adj2).  Such
regions bracket  the   \object{SN\,2008D/XT\,080109} position  in  the
radial direction towards  the  \object{NGC\,2770} nucleus,  both  at a
distance    of             $\sim            10^{\prime\prime}$    from
\object{SN\,2008D/XT\,080109}    (beam     FWHM  at    1.2\,mm   $\sim
11^{\prime\prime}$).    They  were   selected   sufficiently close  to
\object{SN\,2008D/XT\,080109}   to   have   a   good  estimate of  its
background    emission,  but sufficiently far   to  avoid  most of the
\object{SN\,2008D/XT\,080109}  flux  within the   beam.   The adjacent
1.2\,mm pointings yielded fluxes consistent with those obtained at the
\object{SN\,2008D/XT\,080109}   position    (see  Table~\ref{table2}),
implying  that  the  1.2\,mm data   are  very likely dominated  by the
\object{NGC\,2770}      background      flux    and    not          by
\object{SN\,2008D/XT\,080109}.

\begin{table}[t]
\centering
\begin{center}
\caption[]{\label{table2} Log of the millimetre observations.}
\begin{tabular} {cccccc}
\hline
\hline
\hspace{-0.15cm} Date$^a$          &\hspace{-0.35cm}Telescope         & \hspace{-0.3cm} Field& \hspace{-0.4cm} Integration & \hspace{-0.4cm} $\lambda$ \hspace{-0.45cm}& \hspace{-0.45cm}  Flux density \\
\hspace{-0.15cm} 2008, UT          &\hspace{-0.35cm}                  & \hspace{-0.4cm}      & \hspace{-0.4cm} time (min)  & \hspace{-0.4cm}   mm      \hspace{-0.4cm}& \hspace{-0.45cm} [mJy/beam]     \\
\hline
\hline\noalign{\smallskip}
 \hspace{-0.15cm} Jan 25.931       &\hspace{-0.35cm} IRAM 30m& \hspace{-0.3cm}\object{SN\,2008D} & \hspace{-0.4cm} $3 \times 20$ & \hspace{-0.45cm} 1.2 \hspace{-0.45cm}& \hspace{-0.35cm} 2.46 $\pm$ 0.54 \\
 \hspace{-0.15cm} Jan 28.200       &\hspace{-0.35cm} IRAM 30m& \hspace{-0.3cm}\object{SN\,2008D} & \hspace{-0.4cm} $1 \times 20$ & \hspace{-0.45cm} 1.2 \hspace{-0.45cm}& \hspace{-0.35cm} 2.29 $\pm$ 1.20 \\
 \hspace{-0.15cm} Jan 30.171       &\hspace{-0.35cm} IRAM 30m& \hspace{-0.3cm}\object{SN\,2008D} & \hspace{-0.4cm} $2 \times 20$ & \hspace{-0.45cm} 1.2 \hspace{-0.45cm}& \hspace{-0.35cm} 1.44 $\pm$ 0.60 \\
 \hspace{-0.15cm} Jan 25.931-30.171&\hspace{-0.35cm} IRAM 30m& \hspace{-0.3cm}\object{SN\,2008D} & \hspace{-0.4cm} $6 \times 20$ & \hspace{-0.45cm} 1.2 \hspace{-0.45cm}& \hspace{-0.35cm} 2.03 $\pm$ 0.38$^b$\\
 \hspace{-0.15cm} Jan 31.925       &\hspace{-0.35cm} IRAM 30m& \hspace{-0.3cm}Adj1$^{c}$         & \hspace{-0.4cm} $2 \times 16$ & \hspace{-0.45cm} 1.2 \hspace{-0.45cm}& \hspace{-0.35cm} 2.39 $\pm$ 0.70 \\
 \hspace{-0.15cm} Jan 31.936       &\hspace{-0.35cm} IRAM 30m& \hspace{-0.3cm}Adj2$^{d}$         & \hspace{-0.4cm} $2 \times 16$ & \hspace{-0.45cm} 1.2 \hspace{-0.45cm}& \hspace{-0.35cm} 2.81 $\pm$ 0.71 \\
                                \hline
 \hspace{-0.15cm} Jan 23.854       &\hspace{-0.35cm}   PdB & \hspace{-0.3cm}\object{SN\,2008D}   & \hspace{-0.4cm} $1 \times 60$ & \hspace{-0.45cm} 3.3 \hspace{-0.45cm}& \hspace{-0.35cm} 0.65 $\pm$ 0.15 \\
 \hspace{-0.15cm} Nov 9.336        &\hspace{-0.35cm}   PdB & \hspace{-0.3cm}\object{SN\,2008D}   & \hspace{-0.4cm} $1 \times 111$& \hspace{-0.45cm} 2.9 \hspace{-0.45cm}& \hspace{-0.35cm} 0.03 $\pm$ 0.10 \\
\noalign{\smallskip}
\hline
 \multicolumn{6}{l}{~\scriptsize{$a$~~~ Mean observing epoch.}}\\
 \multicolumn{6}{l}{~\scriptsize{$b$~~~ Weighted average of the three previous table lines.}}\\
 \multicolumn{6}{l}{~\scriptsize{$c$~~~ R.A.$=$09$^{\rm h}$~09$^{\rm m}$~30$\fs$085, Dec.$=$+33$^{\circ}$~08$'$~29$\farcs$98 (J2000).}}\\
 \multicolumn{6}{l}{~\scriptsize{$d$~~~ R.A.$=$09$^{\rm h}$~09$^{\rm m}$~31$\fs$211, Dec.$=$+33$^{\circ}$~08$'$~10$\farcs$25 (J2000).}}\\
\hline
\end{tabular}
\end{center}
\end{table}

In addition,  a $4.3\sigma$ detection was   achieved on January  23 at
3.3\,mm with  the  Plateau de Bure   (PdB)  interferometer centered on
\object{SN\,2008D/XT\,080109}   (see Table~\ref{table2})    with    an
observing beam size   of   $2\farcs75 \times 1\farcs11$.     The  last
millimetre observation was  carried out on November  9.336 UT with PdB
at 2.9\,mm, yielding  a 3$\sigma$ upper limit of  0.30 mJy/beam.  This
non-detection suggests that the flux detected at 3.3\,mm on January 23
mostly originated from  \object{SN\,2008D/XT\,080109} and not from the
host galaxy dust.

\section{Results}

\subsection{Properties of the polarizing ISM towards \object{SN\,2008D/XT\,080109}}

The Galactic    reddening  towards   \object{NGC\,2770}  is  very  low
(Schlegel et al. 1998, $E(B-V)=0.032$), implying a $V$-band extinction
of    only $A_{V}=0.1$ mag  (Cardelli   et  al. 1989).   Therefore the
contribution of  the Galactic dust  to the   measured 1.2\,mm flux  is
negligible.  Thus,  assuming that most of  the 1.2\,mm  emission comes
from  optically  thin dust emission  in  the  host galaxy interstellar
medium (ISM) towards  \object{SN\,2008D/XT\,080109}, it is possible to
roughly estimate the host optical extinction as:  ${A_{V}} = (1.086 \,
S_{1.2})/(B_{1.2}(T_{\rm dust})  \,  \Omega_{mb}) \times (\kappa_{V} /
\kappa_{1.2})$.  $B_{1.2}(T_{\rm dust})$ is the Planck function of the
dust at a temperature  $T_{\rm dust}$, $\Omega_{mb}$ is the  main-beam
solid  angle, $S_{1.2}$ is the flux  density per  beam at 1.2\,mm, and
$\kappa_{V}/\kappa_{1.2}$ is  the   ratio of  the visual    extinction
coefficient to  the  1.2\,mm dust opacity, which  was  estimated to be
$\langle           \kappa_{V}/\kappa_{1.2}\rangle=(4\pm2)\times10^{4}$
(Kramer et al. 1998).

In order to  estimate  $A_{V}$ a value of   $T_{\rm dust}$ has to   be
assumed.  Domingue et al. (1999) inferred $T_{\rm dust}=21\pm2$\,K for
the colder dust component of a set of three spiral galaxies similar to
\object{NGC\,2770}.    On   the   other   hand, the    spectral energy
distribution of \object{NGC\,2770}  integrated over the entire  galaxy
yields $T_{dust}=30\pm5$\,K (Th\"one et al. 2009).  So we adopted both
$T_{dust}=21\pm2$\,K and $T_{dust}=30\pm5$\,K.

Using    $S_{1.2}=(2.03\pm0.38)$\,mJy/beam    (line         four    of
Table~\ref{table2}; weighted average of  the first three  table lines)
and  both  $T_{\rm  dust}=21\pm2$\,K and $T_{\rm  dust}=30\pm5$\,K, we
obtain  $A_{V}=0.43\pm  0.31$ and  $A_{V}=0.27\pm0.15$,  respectively.
However,  as $S_{1.2}$ corresponds to the  flux integrated in the beam
along the  projected thickness   of  the host galaxy   (which includes
contribution from any ISM located beyond  the XT), we can only provide
an upper limit  to $A_{V}$.   To be conservative,  we will  assume  an
extinction of       $A_{V}<0.43+0.31=0.74$,   obtained  with   $T_{\rm
  dust}=21\pm2$\,K.  This $A_{V}$ value is $\sim  0.5 - 1.8$ mag lower
than the line-of-sight $A_{V}$ values inferred  by other authors based
on a broad diversity of techniques (Mazzali  et al. 2008; Soderberg et
al. 2008;  Malesani et al. 2009;  Modjaz et al. 2009).  An unrealistic
temperature of $T_{\rm dust}\sim10$\, K would be  necessary to have an
$A_{V}$ value in agreement with the above authors.

This apparent disagreement can   be explained by the  impossibility to
resolve a   clumpy  host    galaxy   ISM  with   our   1.2\,mm    beam
(FWHM=$11^{\prime  \prime}$).   The clumpy ISM   would show three main
properties;  $i)$ an  ISM composed  of cells  superimposed on a  lower
extinction  background, $ii)$  cells displaying typical  angular sizes
smaller than the beam size,  $iii)$ typical angular separation between
cells small enough to allow the beam to contain  several cells.  Given
that the extinction determined  by the 1.2\,mm data  is an  average of
the integrated flux received in  the beam, this scenario would explain
the   higher extinction  derived  along  the  XT line-of-sight  (i.e.,
spectroscopy).  On the other  hand, condition $iii)$ would explain the
fact that the  pointings adjacent to the \object{SN\,2008D/XT\,080109}
position yielded similar 1.2\,mm fluxes, and hence extinction values.

We  can estimate a  rough upper  limit of   the typical cell  sizes as
follows.   First, we assume that   the flux  differences (or  internal
dispersion)  between   our      three  1.2\, mm       pointings   (see
Table~\ref{table2};        weighted      flux      average          of
\object{SN\,2008D/XT\,080109}, Adj1 and Adj2) were entirely due to the
statistical fluctuations    in  the average   number   of cells  ($N$)
contained in  each beam.  In   our case the   three 1.2\, mm pointings
yield $2.03\pm0.38$, $2.39\pm0.70$ and  $2.81\pm0.71$ mJy/beam, so the
dispersion is 0.41 mJy/beam.  If the cells were identical and randomly
distributed, then  the  ratio between    the  average flux  and    the
dispersion of the 3  pointings would be  approximately $\sqrt{N}$.  In
our case the average flux of the 3  pointings (calculated by weighting
with the corresponding  flux errors) is 2.24  mJy/beam, so this yields
$N  \sim 30$ cells/beam.  However,  the dispersion is  very likely not
only due   to  statistical fluctuations of   $N$,  i.e, the dispersion
probably  also has an instrumental/calibration/photon-noise component.
So the statistical fluctuations due to $N$ would  be likely lower than
0.41 mJy/beam and therefore we can only set a lower limit of $N > 30$.

\begin{figure}[t]
\begin{center}
     \vspace{0.2cm}
     \includegraphics[width=9.4cm]{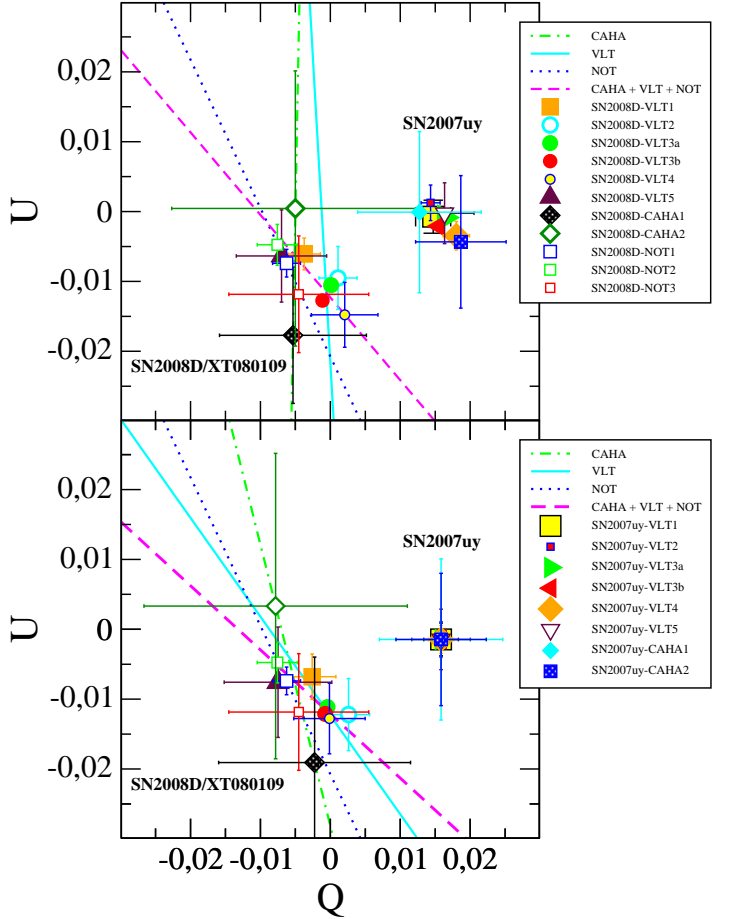}
     \caption{\label{Fig1}  {\em  Upper  panel:} Stokes parametres  of
       \object{SN\,2008D/XT\,080109} and \object{SN\,2007uy}, once the
       GIP correction has  been  included.  The cloud of  measurements
       representing \object{SN\,2007uy} seems  to be more compact than
       the \object{SN\,2008D/XT\,080109}  one.    The  different lines
       represent  the linear fits to the \object{SN\,2008D/XT\,080109}
       Stokes  parametres obtained  when   the CAHA  (dot-dashed), VLT
       (continuous), NOT  (dotted)  and  all the  data  (VLT+CAHA+NOT,
       long-dashed)  are  considered.    The directions   of all   the
       straight lines       are    statistically    consistent    (see
       Table~\ref{Table4}),  which  might indicate the  presence  of a
       preferential axis.  This possible evidence is strengthened when
       \object{SN\,2007uy}  is used  as reference   (see below).  {\em
         Bottom       panel:}      Stokes         parametres        of
       \object{SN\,2008D/XT\,080109}   and   \object{SN\,2007uy},   if
       \object{SN\,2007uy}  is  kept fixed  at   its barycentre.   The
       directions  of all  the  straight lines are  statistically more
       consistent than    in the upper    panel (compare  the internal
       consistency of the different orientations in Table~\ref{Table3}
       and  Table~\ref{Table4}),  reinforcing the possible presence of
       an instrumental-independent dominant  symmetry axis present  in
       the \object{SN\,2008D/XT\,080109} data.}
\end{center}
\end{figure}

\begin{figure*}[t]
\begin{center}
      \includegraphics[width=16cm]{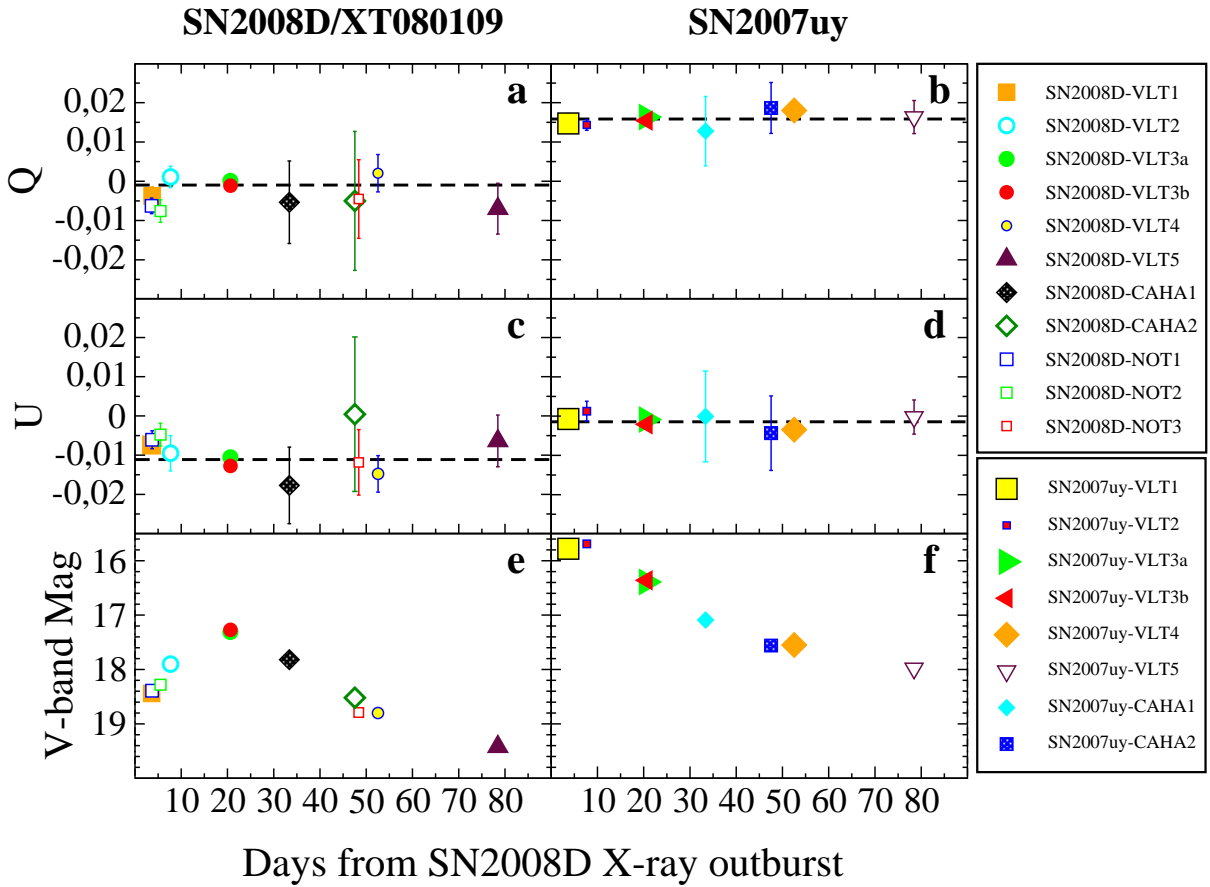}
     \caption{\label{Fig3}  Polarimetric and photometric evolution
         of   \object{SN\,2008D/XT\,080109} (panels  a,c,e)        and
         \object{SN\,2007uy}  (panels b,d,f).  For consistency we used
         the same symbols as in Fig.~\ref{Fig1}. The long-dashed lines
         indicate the position of the barycentre ($\langle Q \rangle$,
         $\langle U \rangle$)  of each SN.   The two bottom  panels (e
         and      f)  show       the   $V$-band   light    curves   of
         \object{SN\,2008D/XT\,080109}       and   \object{SN\,2007uy}
         uncorrected for Galactic reddening.  None of the two SNe seem
         to obey any  obvious smooth  long-term polarimetric evolution
         correlated with their optical light curves.  }
\end{center}
\end{figure*}

Additionally, in order to match the $A_{V}$ derived  from the 1.2\, mm
flux with the spectroscopic $A_{V}$ values reported in the literature,
the emitting region (the  total area covered  by the cells  within the
beam  solid angle)   should fill only\footnote{For   dust temperatures
  ranging from $T=21$ to $T=30$\,K.}  10-30\%  of $\Omega_{mb}$.  This
would explain the  low $A_{V} < 0.74$  mag derived using the 1.2\,  mm
flux  (Kramer    et al.  2008)  for    cells  having  a  line-of-sight
spectroscopic  extinction of   $A_{V}=1.2-2.5$  mags.  Thus,  assuming
identical circular  cells,   a \object{NGC\,2770} distance   of 27 Mpc
(Soderberg et al. 2008),  and a $\Omega_{mb}$  filling factor of 30\%,
we can impose a maximum  cell diameter of  $D < 0.16$ kpc.  Adopting a
beam  filling factor of   10\%, we would  obtain  smaller cells, so we
consider 0.16  kpc as  a robust upper   limit.  As a comparison  it is
interesting  to mention that  most ($\sim98$\%) of the giant molecular
clouds resolved  by millimetric observations  in the nearby spiral M33
show  sizes between  0.05  and 0.16  kpc  (Rosolowsky  et al.   2007),
compatible with our observations.   Similarly,  most of the  H{\rm II}
regions in nearby spiral galaxies show physical scales consistent with
$<0.16$  kpc (Gonz\'alez-Delgado \& P\'erez 1997).    We note that the
dominant $P$  is the result  of all the  material integrated along the
line-of-sight,   so we can   not conclude   that  the dominant  $P$ is
exclusively due to the XT circumstellar dust.

\subsection{Properties  of \object{SN\,2008D/XT\,080109} and \object{SN\,2007uy} on the Stokes plane}
\label{polaanalysis}

The  $P$  and  $\theta$ values   displayed  in Table~\ref{table1} were
corrected for the GIP and the polarimetric statistical bias (Wardle \&
Kronberg 1974).  The further polarimetric  analysis will deal with the
properties of $Q$ and $U$ of both objects  on the Stokes plane, so the
$P$ bias correction factor  (included in Table~\ref{table1})  will not
be considered.

\subsubsection{Is \object{SN\,2008D/XT\,080109} polarimetrically variable?}
\label{variability}

The    Stokes    parametres    of   \object{SN\,2008D/XT\,080109}  and
\object{SN\,2007uy},  corrected    for  the   GIP,  are   plotted   in
Fig.~\ref{Fig1}.  Both  events show polarization levels  of $1-1.5$\%,
significantly       lower   than     the      one    measured       in
\object{XRF\,060218/SN\,2006aj} (Gorosabel et  al.  2006; Maund et al.
2007).   Assuming both  a  GIP  law  and   a Galactic  extinction  law
(Serkowski et al. 1975; Cardelli et  al. 1989) for \object{NGC\,2770},
the $A_{V}$ values  reported in the literature ($\sim  1.2 - 2.5$ mag)
would yield a conservative HGIP polarization upper limit of $P_{max} =
2.9\%  \times A_V \sim 3.5 -  7.3$\%, consistent with the polarization
measured for \object{SN\,2008D/XT\,080109} and \object{SN\,2007uy}.

As seen in the upper  panel  of Fig.~\ref{Fig1},  the cloud of  points
corresponding to \object{SN\,2007uy}  seems to be more clustered  than
the one of  \object{SN\,2008D/XT\,080109}.     For the  two    stellar
explosions   we quantified the probability  that  the Stokes parametre
distributions  are consistent  with  no   time evolution.  First,   we
double-checked using Monte Carlo  methods that, for the average number
of   counts  of   our images,    both $Q$   and   $U$  follow Gaussian
distributions.    This  fact  assures  that  the   $\chi^2$ test is an
appropriate statistical tool.  Second, we determined the barycentre of
the two  objects on the Stokes plane  ($\langle Q \rangle$, $\langle U
\rangle$;  see the long-dashed  lines   of Fig.~\ref{Fig3}).  Then  we
calculated for  the $Q$  and $U$ of   the two events the  $\chi^2/$dof
value with respect  to $\langle Q   \rangle$ and $\langle U  \rangle$,
respectively\footnote{$\chi_Q^2/{\rm  dof} = \sum_{i=0}^k (\frac{Q_i -
    \langle Q \rangle}{\Delta\,Q_i})^2  ~/~  (k-1)$ and $\chi_U^2/{\rm
    dof}     =      \sum_{i=0}^k    (\frac{U_i    -        \langle   U
    \rangle}{\Delta\,U_i})^2 ~/~ (k-1)$,  $k$   being the number    of
  visits and dof=$k-1$ the degrees of freedom.}.

For   \object{SN\,2007uy}   the   distribution  of   $Q$  ($U$)  shows
$\chi^2/$dof=$  4.18/7$   ($\chi^2/$dof=$   4.32/7$)  with  respect to
$\langle      Q     \rangle$   ($\langle     U     \rangle$).      For
\object{SN\,2008D/XT\,080109}  the  distribution  of $Q$ ($U$) yields
$\chi^2/$dof=$ 19.06/10$  ($\chi^2/$dof=$  22.96/10$), clearly  larger
than  the $\chi^2/$dof value of  \object{SN\,2007uy}.  These values of
$\chi^2/$dof were used to obtain the corresponding probabilities.  For
\object{SN\,2007uy} the probability that  $Q$ and $U$ are constant are
0.758  and 0.742, respectively.  \object{SN\,2008D/XT\,080109}  shows
lower  probabilities,  0.040    and 0.011,  respectively.    Thus  the
probabilities that  $Q$ and $U$  are simultaneously constant are 0.563
for       \object{SN\,2007uy}   and       $4.3\times10^{-4}$       for
\object{SN\,2008D/XT\,080109}. 

We  are aware that different error  sources affecting the measurements
(background  determination   uncertainties, flat     field  correction
inaccuracies, no instrumental  polarization correction available along
the FoV for the CAHA and NOT images, etc.) might still not be included
in  the   error  bars,  so   the  values  of  $\chi^2/$dof   could  be
overestimated for  both  sources  (and hence  the  above probabilities
underestimated).     Thus, we  repeated   the  exercise augmenting the
photometric  error bars  of both  objects and  all the  field stars by
20\%.   If we do  that,  the probability that both    $Q$ and $U$  are
constant   are   0.791 for  \object{SN\,2007uy}  and    0.021 for 
  \object{SN\,2008D/XT\,080109}, respectively.

  We have  checked, if due to some  uncontrolled reason, the ($Q$,$U$)
  distribution on  the Stokes plane is just  the result of mixing data
  acquired with different telescopes/instruments.  In order to inspect
  the existence of this potential instrumental  artifact, we redid the
  statistical analysis separating the images that contain both objects
  according to the telescope used (hence the NOT is excluded from this
  comparative   study       between         \object{SN\,2007uy}    and
  \object{SN\,2008D/XT\,080109}   since  \object{SN\,2007uy}  was  not
  imaged by the  NOT).  Thus we determined  separately for the VLT and
  CAHA the probabilities of having constant Stokes parametres.

  Using  only the VLT data  points, we determined that the probability
  that both $Q$ and $U$ are constant are 0.295 for \object{SN\,2007uy}
  and 0.007 for  \object{SN\,2008D/XT\,080109},  respectively.  If  we
  enlarge the  photometric errors by 20\%  the  probabilities would be
  then     0.532     and     0.056,    for  \object{SN\,2007uy}    and
  \object{SN\,2008D/XT\,080109},     respectively.         Considering
  exclusively the  CAHA data-set we can   not reach strong conclusions
  given  the  reduced number  of  visits   (2)  and large error  bars.
  However, even with the  limited CAHA data, \object{SN\,2007uy} shows
  a higher probability   than  \object{SN\,2008D/XT\,080109} of  being
  polarimetrically    constant,  0.457  versus    0.403.   If the CAHA
  photometric errors are augmented by 20\%, then the probabilities are
  0.530      and        0.423      for    \object{SN\,2007uy}      and
  \object{SN\,2008D/XT\,080109}, respectively.

  Independent   of the   data   subset considered   and the additional
  photometric  uncertainty  introduced,  \object{SN\,2008D/XT\,080109}
  reiteratively displays lower probabilities than  \object{SN\,2007uy}
  of being  polarimetrically constant.  In fact, it  is not obvious to
  explain    how  non-accounted   photometric/calibration/instrumental
  uncertainties could systematically   affect only the values  derived
  for    \object{SN\,2008D/XT\,080109}      and   not        those  of
  \object{SN\,2007uy},  which  consistently shows  a higher  degree of
  clustering on the Stokes plane.  It is interesting to note that this
  is against     the  a priori expectations,     given  that  the more
  inhomogeneous host background  of \object{SN\,2007uy} (in comparison
  to     \object{SN\,2008D/XT\,080109})  might  represent  a potential
  uncontrolled  source  of noise  in its  Stokes parametres.   All the
  above arguments  seem to  suggest that \object{SN\,2008D/XT\,080109}
  could show an intrinsic variable polarization component added to the
  dominant  HGIP.      This      suggestion  is    reinforced     when
  \object{SN\,2007uy} is used  as  a secondary  polarimetric  standard
  (see Sect.~\ref{axis}).   However, it is  important  to note that  a
  constant polarization signal from \object{SN\,2007uy} is atypical of
  previously  observed core-collapse SNe   (Wang \& Wheeler 2008).  We
  point out that none of  the two SNe seem  to obey any obvious smooth
  long-term polarimetric  evolution  correlated to  their light curves
  (see Fig.~\ref{Fig3}).

\subsubsection{A symmetry axis for \object{SN\,2008D/XT\,080109}?}
\label{axis}

Given that both \object{SN\,2007uy} and \object{SN\,2008D/XT\,080109}
were  imaged simultaneously  with     VLT   and CAHA,  we    can   use
\object{SN\,2007uy}    as  a  reference   star   with  constant Stokes
parametres.  In  other words,  we can  assume, justified by  the above
probabilities, a  constant projected  geometry for \object{SN\,2007uy}
plus a   constant  HGIP.  As    an additional   argument,  it is  also
interesting to  note   that   the agreement    between   the synthetic
broad-band points reported by Maund et al. (2009) and our data is even
better when \object{SN\,2007uy} is fixed on the Stokes plane.

\begin{figure}[t]
\begin{center}
    \includegraphics[angle=-90,width=\hsize]{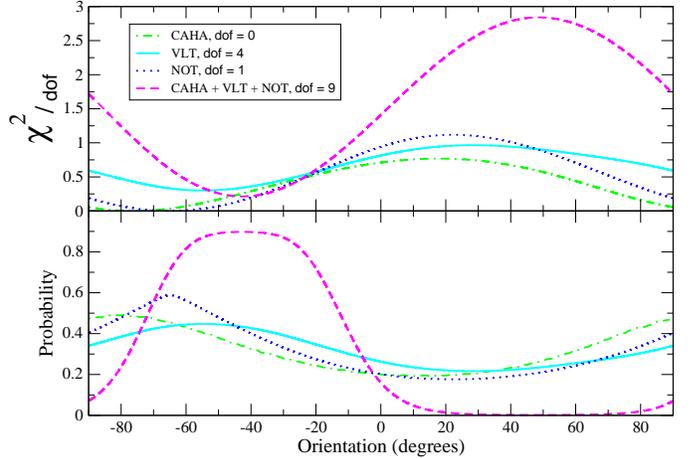}
    \caption{\label{Fig2} {\em   Upper  panel:} The   evolution of the
      linear  fit $\chi^2/$dof  as a   function   of the fitted   line
      orientation   when constant    polarization   is   assumed   for
      \object{SN\,2007uy}.   The  minima   correspond  to the  optimal
      orientation  of    the  straight    lines  displayed   in    the
      Fig.~\ref{Fig1} bottom  panel, also given in Table~\ref{Table3}.
      As   seen the  global   shape of  the   evolution for  the three
      independent data-sets (VLT, NOT and  CAHA) are quite equivalent,
      their minima  being  systematically located  at similar negative
      angles.  The  long-dashed line shows  the $\chi^2/$dof evolution
      when all  the data are jointly considered.   We note that in the
      case  of  CAHA,  given   that  dof=$0$,  $\chi^2/$dof   actually
      represents $\chi^2$.  {\em  Bottom panel:} The evolution  of the
      normalized probability  associated  with the   above panel.   As
      displayed the three independent data-sets show their probability
      maxima at a   comparable  orientation  angle.  The maximum    is
      reinforced when all   the data are  combined (long-dashed line).
      Given that   dof=$0$ for   CAHA, the   corresponding probability
      distribution was obtained through   a  Monte Carlo  method.   We
      refer the reader to the main text for further information.}
\end{center}
\end{figure}

Thus, using  \object{SN\,2007uy} as  calibrator, a  linear  fit to the
\object{SN\,2008D/XT\,080109} data  yields        a       satisfactory
$\chi^2/$dof=$1.9/9$ (long-dashed  straight line   of  Fig.~\ref{Fig1}
bottom panel).    Hence,    the  \object{SN\,2008D/XT\,080109}  Stokes
parametres might suggest the  existence  of a dominant symmetry  axis,
offset from the origin  due to the  HGIP.  Furthermore, as seen in the
first  three lines of  Table~\ref{Table3}  the inferred symmetry  axes
show  a  preference  to   negative   orientations independent of   the
telescope+instrument  employed in the observations\footnote{We  define
  the  orientation as   the angle of   the fitted  straight line  with
  respect to the horizontal axis representing $Q$ on the Stokes plane.
  The orientation ranges between -90  and +90 degrees.}.  This  effect
can be seen  in the bottom panel  of Fig.~\ref{Fig1},  which shows the
linear fits carried out both  when the data are considered  separately
(CAHA, VLT,   NOT) and jointly  (CAHA+VLT+NOT).  The  linear fits were
obtained minimizing  $\chi^2$,  which is    defined as the    weighted
distance perpendicular to the fitted  line\footnote{Also known as {\it
    weighted total least squares} (WTLS), {\it weighted rigorous least
    squares},  or  {\it weighted  orthogonal  regression  method} (see
  Lemmerling  \&  van Huffel  2002,    and references therein).}    We
considered this $\chi^2$ definition because the variables to be fitted
($Q$,$U$) and  their corresponding  errors  are treated  symmetrically
(Boogs et al. 1990; Babu \& Feigelson 1996).

\begin{table*}[t]
\centering
\begin{center}
  \caption[]{\label{Table3}  Linear  fits   on  the  Stokes plane   of
    \object{SN\,2008D/XT\,080109}.  The  linear fits are  shown in the
    bottom panel of Fig.~\ref{Fig1}, where we used \object{SN\,2007uy}
    as a  reference star  with constant  Stokes  parametres.  Column 1
    displays the telescope  used and column  2 the number of visits or
    points on the Stokes plane.  Column 3 provides the $\chi^2/$dof of
    the optimal linear fit.  Column  4 shows the orientation angles at
    the minima (maxima)  of the Fig.~\ref{Fig2} upper  (bottom) panel.
    Columns 5 and 6 provide the lower  and upper error bars around the
    orientation    angle   of the  linear  fit.      As seen the three
    independent  data-sets (CAHA, NOT and  VLT) show orientations well
    consistent within  errors.   Last table  line reports the  results
    when all the data are combined.}
\begin{tabular} {lccccc}
\hline
\hline
Telescope & Number of& Minimum     & Orientation angle & Lower Error Bar & Upper Error Bar \\
          & visits   & $\chi^2/$dof&     (degrees)     &  (degrees)      & (degrees)       \\
\hline
\hline
CAHA      &    2    &  0$^{a}$     & -76.1             & 49.4$^{b}$& 48.5$^{b}$ \\
NOT       &    3    &  0.001/1     & -64.8             & 46.1 & 46.7 \\
VLT       &    6    &  1.198/4     & -54.8             & 52.2 & 51.8 \\
\hline
CAHA+VLT+NOT& 11    &  1.922/9     & -42.6             & 23.3 & 22.9 \\
\hline
\multicolumn{6}{l}{\scriptsize{$a$~~In this particular case, $\chi^2=0$ and dof=$0$ for the minimum, so $\chi^2/$dof could also be formally represented by $0/0$. }}\\
\multicolumn{6}{l}{\scriptsize{$b$~~Errors obtained through Monte Carlo methods.}}\\
\hline
\end{tabular} 
\end{center}
\end{table*}

In order to quantify the degree  of consistency that the symmetry axes
of  the  3 data-sets  could   show (when constant  \object{SN\,2007uy}
Stokes parametres are assumed)  we mapped the  evolution of the linear
fit $\chi^2/$dof  when the orientation of the  symmetry axis is varied
from $-90$ to $+90$ degrees.  The upper panel of Fig.~\ref{Fig2} shows
this evolution.  We  warn the reader that  in  the particular  case of
CAHA dof=$0$, so what is plotted for CAHA in the Fig.~\ref{Fig2} upper
panel  actually represents  $\chi^2$.   As seen  all  the three
independent data-sets show  their $\chi^2/$dof minima approximately at
the same orientation.   The  position of all the  $\chi^2/$dof  minima
were positively confirmed  using the {\tt  fitexy} routine implemented
by Press et al.  (1992).  The lower panel of Fig.~\ref{Fig2} shows the
normalized probabilities corresponding  to the $\chi^2/$dof values  of
the above panel.

For the  case of  CAHA, given that  dof=$0$,  the  plotted probability
evolution    was obtained by  simulating   mock CAHA Stokes parametres
created by a weighted Monte Carlo method  assuming Gaussian errors for
$Q$   and  $U$.  Then given  two   simulated  ($Q,U$)  CAHA  pairs the
orientation  of  the line passing through   those Stokes parametres is
determined.  The repetition of  this  process allowed us to  construct
the histogram of  the line orientations for  the CAHA data.  Thus, the
lower panel  of  Fig.~\ref{Fig2}  displays for CAHA  the  histogram of
orientations for $10^{7}$  Monte Carlo simulations\footnote{This Monte
  Carlo method, used  for determining the probability distribution for
  CAHA, is not optimal for the VLT and NOT data.   For the VLT and the
  NOT a set  of simulated Stokes parametres does  not determine a line
  passing through them (they contain more than  2 visits) and a formal
  fit  is necessary.  So,   the  histogram of orientations  would  mix
  linear fits with different qualities.}.

For each  data-set the upper  and lower error  bars in the orientation
angle  (displayed   in  Table~\ref{Table3}, columns    5  and 6)  were
calculated by  integrating   the  $\pm34.145$\%   percentile  of   the
probability distribution around the corresponding maximum.  As seen in
Table~\ref{Table3} the  axes  inferred   from the  three   independent
data-sets are consistent within  errors.  Hence, the inferred symmetry
axes are  statistically consistent  in  all  cases, regardless of  the
instrumental setup  used to determine  it.  It  is also interesting to
remark   that  the   NOT   Stokes   parametres, which    did   not use
\object{SN\,2007uy}  as a secondary  calibrator (so they might apply a
slightly different GIP correction), are  also satisfactorily fitted by
a straight line.   Furthermore, the orientation of  the NOT linear fit
agrees  with the ones inferred  based on the  VLT  and CAHA data.  All
these arguments strengthen the reality of this possible symmetry axis.

The  existence of  a  symmetry axis   on  the  Stokes plane could   be
explained by an  axisymmetric   explosion where the direction   of the
symmetry axis is constant, but the eccentricity  evolves with time. We
stress that  our polarimetric data  can  provide information about the
projected geometry on the plane of the sky  exclusively.  Thus, we can
not discard that the constant Stokes parametres of \object{SN\,2007uy}
could be caused by a variable eccentricity expansion when it is viewed
pole-on.

         \subsubsection{The   robustness     of     the   possible
      \object{SN\,2008D/XT\,080109} symmetry  axis: must  we assume a
      constant geometry for \object{SN\,2007uy}?}
    \label{robust}
    One     might  wonder    if  the    possible      symmetry axis of
    \object{SN\,2008D/XT\,080109} is    just an artifact due   to the
    assumption of a  constant geometry  for \object{SN\,2007uy}.  Thus
    we repeated the exercise  explained in Sect.~\ref{axis} using  the
    original   unrecalibrated \object{SN\,2008D/XT\,080109}    Stokes
    parametres.    The results are  displayed  in  the upper panel  of
    Fig.~\ref{Fig1}   and in  Table~\ref{Table4}.   Note  that the NOT
    Stokes  parametres, could not  be   recalibrated,  so the  fit  is
    identical to the one displayed in Table~\ref{Table3}.

    The \object{SN\,2008D/XT\,080109}  entire  data set (CAHA+VLT+NOT)
    is still  satisfactorily fitted ($\chi^2/dof$=1.066) by a straight
    line.  The  derived orientation ($-49.8^{+17.2}_{-17.4}$  deg), is
    fully  consistent with  the  orientation  determined based on  the
    recalibrated  Stokes parametres ($-42.6^{+22.9}_{-23.3}$ deg).  On
    the other  hand, the three  separated data sets yield orientations
    with   internal  consistency  within     1$\sigma$,  so   as    in
    Sect.~\ref{axis}   the   inferred   orientations  seem   to     be
    instrumentally independent.     It is   also noticeable that   the
    recalibrated Stokes parametres  yield orientations (-76.1,  -64.8,
    and   -54.8   deg)  with   lower   internal dispersion   than  the
    unrecalibrated ones (88.9,-64.8,-86.7 deg).

    The    fit  quality to   the   VLT  Stokes   parametres is clearly
    reduced with respect  to  the  recalibrated case, but  it  is
    still fairly satisfactory  ($\chi^2/dof=1.375$).  It is worthwhile
    to note that the major source of this $\chi^2/dof$ increment comes
    from    the  lack  of    overlap     within   1$\sigma$ of     the
    \object{SN\,2008D/XT\,080109}  Stokes parametres corresponding to
    VLT3a  and VLT3b (red   and  green circles of the  Fig.~\ref{Fig1}
    upper  panel),  which   appear  slightly separated  (by   $|\Delta
    Q|$=0.0012 and $|\Delta   U|$=0.0022, corresponding to a  relative
    shift of 1.4$\sigma$ and 2.5$\sigma$, respectively).

    One could be  tempted to speculate  on a possible rapid  ($Q$,$U$)
    variability        in    visits      VLT3a      and    VLT3b    of
    \object{SN\,2008D/XT\,080109}.  However, firstly the  polarimetric
    variation  between   VLT3a  and    VLT3b  is not     statistically
    significant, and  secondly (and most  importantly), this  internal
    shift between VLT3a and VLT3b is  essentially mimicked in the same
    visits by  the  behaviour of  \object{SN\,2007uy}.  This  suggests
    that this systematic shift present in both  SNe does not reflect a
    real polarimetric  variation,   but is probably   just  a spurious
    polarization fluctuation due to slightly different GIP corrections
    for  VLT3a and VLT3b  which affects both  SNe in a systematic way.
    In  fact, in   both visits  the  seven stars  (identical  for both
    visits) used for the  GIP are fainter than \object{SN\,2007uy} and
    \object{SN\,2008D/XT\,080109}, so  they show  larger   photometric
    statistical fluctuations.  Thus, when using \object{SN\,2007uy} as
    a  calibrator the  Stokes  shift between  VLT3a  and VLT3b  almost
    disappears,      overlapping             within    $1\sigma$   for
    \object{SN\,2008D/XT\,080109}      (and         obviously      for
    \object{SN\,2007uy}).   This example illustrates the importance of
    using   the   brightest    common     object  in     the     field
    (\object{SN\,2007uy},          even           brighter        than
    \object{SN\,2008D/XT\,080109})  as a  reference  star in  order to
    remove internal spurious shifts  below $\sim0.2$\% from the Stokes
    plane.

    We   conclude that,    at    a  lower significance level,      the
    unrecalibrated data still allow  the existence of a  symmetry axis
    for \object{SN\,2008D/XT\,080109}  to be plausible.  Therefore, in
    order to suggest a symmetry  axis it is  not strictly necessary to
    assume   that   the   \object{SN\,2007uy} geometry  is   constant.
    However, this assumption strengthens its possible existence.

\begin{table*}[t]
\centering
\begin{center}
\caption[]{\label{Table4}   Linear  fits  on   the  Stokes   plane  of
  \object{SN\,2008D/XT\,080109}  when no constant Stokes   parametres
  are assumed for   \object{SN\,2007uy}.   The corresponding  straight
  lines are  shown in  the upper  panel of  Fig.~\ref{Fig1}.  Column 1
  displays the telescope used  and column 2   the number of visits  or
  points  on the Stokes plane.   Column 3 provides the $\chi^2/$dof of
  the optimal linear  fit.  Column 4  shows the orientation angles  at
  the minima (maxima)  of  the Fig.~\ref{Fig2} upper  (bottom)  panel.
  Columns 5  and 6 provide the lower  and upper error  bars around the
  orientation angle of the linear fit.   As seen the three independent
  data-sets (CAHA, NOT and VLT)  show more scattered orientations (and
  also larger  $\chi^2/$dof    values) than  the  ones   displayed  in
  Table~\ref{Table3}, but  they are still consistent  within errors.
  Last table line reports the results when all the data are combined.}
\begin{tabular} {lccccc}
\hline
\hline
Telescope & Number of&    Minimum  & Orientation angle & Lower Error Bar & Upper Error Bar \\
          & visits   & $\chi^2/$dof& (degrees)         & (degrees)       & (degrees)      \\
\hline
\hline
  CAHA    &  2    & 0$^{a}$   &  88.9 & 53.9$^{b}$& 48.8$^{b}$ \\
  NOT     &  3    & 0.001/1   & -64.8 & 46.1      & 46.7 \\
  VLT     &  6    & 5.501/4   & -86.7 & 27.9      & 26.8 \\
\hline
 CAHA+VLT+NOT& 11  & 9.599/9  &-49.8 & 17.4 & 17.2 \\
\hline
\multicolumn{6}{l}{\scriptsize{$a$~~In this particular case, $\chi^2=0$ and dof=$0$ for the minimum, so $\chi^2/$dof could also be formally represented by $0/0$. }}\\
\multicolumn{6}{l}{\scriptsize{$b$~~Errors obtained through Monte Carlo methods.}}\\
\hline
\end{tabular} 
\end{center}
\end{table*}

\subsection{Final remarks}

Maund et al. (2009) report a variation in the position angle of the He
{\rm I}  $\lambda5876$   line based on  the   comparison of their  two
observing  epochs.  Thus, we have   to be cautious  in explaining  the
origin   of  the  possible  \object{SN\,2008D/XT\,080109} polarization
variability.  The \object{SN\,2008D/XT\,080109} polarization variation
suggested   by  the analysis of    the  broad-band data  could  not be
explained, totally or  partially, just  by an axisymmetric  expansion.
We can not discard that the polarization variability could be, to some
degree,   caused   by   the  changing  relative    strengths   in  the
line/continuum emission, only  one   of  which may be  polarized,   as
discussed by Maund et  al.  (2009).   Thus $V$-band polarimetric  data
alone can not  resolve  these potential  components, and in  principle
only  spectropolarimetry  could study  the  impact of  this, and other
possible effects, on the evolution of the Stokes parametres.  However,
the  limited time-coverage of  the  spectropolarimeric data (only  two
epochs   available, covering 2 weeks  around  the light curve maximum)
does not make it possible to derive strong conclusions about potential
effects which could impact our  eleven visits spanning $\sim 75$ days.
We can  only   conclude that  our data suggest   a preference   to  an
approximately axisymmetric configuration,  although we can not exclude
deviations from   axisymmetry  towards  a more complex    geometry, as
discussed by Maund et al. (2009).

Since         \object{SN\,2007uy}        occurred    earlier      than
\object{SN\,2008D/XT\,080109},  we have  checked   to what   degree  a
possible early  fast-evolving epoch   of \object{SN\,2007uy}  may have
been missed by our observations.  An inspection of the light curves of
the two events shows that our data cover the rising phase, the maximum
and the decay   of   both objects.   The polarization   inferred   for
\object{SN\,2007uy}   in these   3   phases is  consistent with  being
constant.    Obviously, we can  not  discard  that \object{SN\,2007uy}
showed polarization variability   earlier than our  first epoch, later
than  the  last epoch  or   in the unavoidable  temporal  gaps between
observations.    So,     we    stress   that   our     conclusions  on
\object{SN\,2007uy}  and \object{SN\,2008D/XT\,080109}  only refer  to
the time covered by our observations.

Under     the    mentioned   limitations,    the     distribution   of
\object{SN\,2007uy} on the Stokes plane imposes an  upper limit on the
variability of its  polarization.   First,  we calculated the   $\pm$1
sigma confidence variability  interval of $Q$  ($U$) around $\langle Q
\rangle$ ($\langle U  \rangle$) using the weighted  variance\footnote{
  As  defined by  the N.I.S.T.  (National  Institute  of Standards and
  Technology;              Gaithersburg,       Maryland,          USA,
  http://www.nist.gov/index.html),     $S_Q    =   \sqrt{\frac{N}{N-1}
    \sum_{i=0}^{N} w_{Q_i}  (Q_i    - \langle  Q  \rangle)^2}$,  being
  $w_{Q_i}=\frac{1/\sigma^{2}_{Q_i}}{\sum_{j=0}^{N}\sigma^{2}_{Q_j}}$,
  $\sigma_{Q_i}$ the standard deviation  of $Q_i$, and $N$  the number
  of  \object{SN\,2007uy}  visits.     The  expression of   $S_U$   is
  analogous,  replacing $Q_i$ with $U_i$,    $\langle Q \rangle$  with
  $\langle U  \rangle$, and $\sigma_{Q_i}$  with $\sigma_{U_i}$.}   of
the $Q$  ($U$) sample as  an statistical estimator.  This estimator is
suitable because both $Q$ and   $U$ follow Gaussian distributions  (as
opposed   to  $P$).  Then,   the  ($Q$,$U$)  weighted  variances  were
propagated yielding  a $\pm$34.145\%   confidence interval  of $\Delta
P_{\pm34.145\%} =  ^{+0.08}_{-0.09}$\% around  the \object{SN\,2007uy}
barycentre.  Finally, using the analytic  expression (A1) given in the
Appendix of Wardle   \&  Kronberg (1974)   we estimated  a   $\pm$45\%
confidence   interval  of  $\Delta  P_{\pm45\%}  =^{+0.14}_{-0.15}$\%.
Thus, we estimate  that  over   the course   of  our observations   of
\object{SN\,2007uy} its total  polarization signal does not  change by
more than  0.29\% (90\% confidence  interval).  We double-checked that
Monte Carlo methods impose  similar limits on  the \object{SN\,2007uy}
polarization variability ($\Delta P_{\pm45\%} \sim 0.3$\%).

\section{Conclusions}

For  both stellar explosions our optical  polarization data seem to be
dominated by the  HGIP.  This conclusion is  supported  by our 1.2\,mm
observations,         performed         at    and      around      the
\object{SN\,2008D/XT\,080109}    position on \object{NGC\,2770}.   The
1.2\,mm measurements are  consistent with  no intrinsic emission  from
\object{SN\,2008D/XT\,080109} at this wavelength, and can be explained
by the host galaxy  dust   emission.  The  $A_{V}$ inferred  from  our
1.2\,mm observations is $\sim$   1.2 mag lower than the  line-of-sight
extinction   values  reported   in the     literature.  This  apparent
contradiction can  be solved if  the \object{NGC\,2770} projected dust
distribution around \object{SN\,2008D/XT\,080109} is composed of dense
clumps ($A_{V} \sim 1.2-2.5$ mag) with a typical size $< 0.16$ kpc and
a low filling factor (10-30\%) in our 1.2\,mm beam.

As a bonus, we also report  the detection of  0.65 mJy/beam at 3.3\,mm
coincident with \object{SN\,2008D/XT\,080109} 14  days after the X-ray
outburst.  This emission  agrees   fairly well  with  the  light curve
modeled at 95 GHz (Soderberg et al. 2008).  Furthermore an observation
at 2.9\,mm carried out  on November 9.3 UT 2008  (304.8 days after the
XT), imposed a 3$\sigma$ flux   upper limit of   0.30 mJy/beam, so  we
conclude that the 3.3\,mm detection  is mostly intrinsic emission from
\object{SN\,2008D/XT\,080109}.

\object{SN\,2008D/XT\,080109} seemed to be  surrounded by an  external
thick He mantle (Mazzali et al. 2008; Modjaz et al. 2009; Soderberg et
al.   2008; Tanaka  et   al.  2009b),   in  contrast  to the   He-poor
\object{XRF\,060218/SN\,2006aj} (Pian  et al.  2006;  Soderberg et al.
2006;  Sollerman et   al.   2006).   Given  that   the asphericity  of
core-collapse supernovae is more pronounced  in the inner layers (Wang
\& Wheeler  2008),  this could  explain  to   some  degree the  modest
\object{SN\,2008D/XT\,080109} intrinsic  polarization in comparison to
\object{XRF\,060218/SN\,2006aj} (Gorosabel et  al. 2006; Maund et  al.
2007).

  Notwithstanding  the important HGIP,   a statistical analysis of the
  distribution of the \object{SN\,2008D/XT\,080109} Stokes  parametres
  suggests     that  \object{SN\,2008D/XT\,080109}  could  exhibit  an
  intrinsic variable polarization  component embedded in  the dominant
  HGIP.  In contrast, the \object{SN\,2007uy} polarization data can be
  described by a  constant   projected geometry.  This   conclusion is
  achieved   independent of   the  data being    considered jointly or
  separately according to the telescope used.

  Assuming that the \object{SN\,2007uy} projected geometry is constant
  in  time,  then the  evolution  of the \object{SN\,2008D/XT\,080109}
  Stokes parametres could  show a symmetry  axis.  The same conclusion
  is   achieved if  no      constant  polarization is    assumed   for
  \object{SN\,2007uy},  but  at  a  lower  significance   level.  This
  possible symmetry axis   is also independent of  the instrumentation
  used to perform   the  observations, supporting its reality.    This
  potential axis could be  approximately explained by  an axisymmetric
  aspherical  expansion  with   variable  eccentricity.  However,  our
  broad-band polarimetric data can not  exclude more complex geometric
  configurations, like   possible  deviations from   axisymmetry.  Our
  results  are consistent    with  the   \object{SN\,2008D/XT\,080109}
  asphericity inferred  from  spectroscopic data (Modjaz et  al. 2009;
  Tanaka et  al. 2009a).  We suggest  that at least  the projected, if
  not the  intrinsic, geometry of  the  two explosive events could  be
  different.

\begin{acknowledgements}
  The research   of JG, AJCT,  MJ and  IA is supported  by the Spanish
  programmes            ESP2005-07714-C03-03,           AYA2007-63677,
  AYA2007-67627-C03-03,    AYA2008-03467/ESP and AYA2009-14000-C03-01.
  This  work has been  partially funded by the Spanish-German Acci\'on
  Integrada HA2007-0079.   AdUP  acknowledges  support  from an    ESO
  fellowship.  AR and  SK acknowledge support by  DFG Kl 766/11-3,  PF
  and  DAK  by  the   Th\"uringer Landessternwarte.   IA  acknowledges
  support by    an  I3P  contract    with the  Consejo    Superior  de
  Investigaciones  Cient\'{\i}ficas    (CSIC).   We thank M.A.~P\'erez
  Torres, C.  Barcel\'o,  D.   Malesani, and L.M.~Sarro  for  fruitful
  comments.  The Dark  Cosmology Centre is  funded by the  DNRF.  This
  paper is based   on  data from the  IRAM   30m telescope, the    PdB
  interferometer, the  NOT, the  VLT and  the  2.2m Telescope of Calar
  Alto.  ALFOSC  is     owned by  Instituto de  Astrof\'{\i}sica    de
  Andaluc\'{\i}a (IAA) and operated at the NOT under agreement between
  the IAA  and NBIfAFG.  We  thank  the IRAM Director  and C.~Thum for
  providing  discretionary observing time at  the  IRAM 30m telescope.
  The German-Spanish  Astronomical  Center,  Calar  Alto,   is jointly
  operated by the MPIA Heidelberg  and IAA (CSIC).  IRAM is  supported
  by INSU/CNRS (France), MPG (Germany)  and IGN (Spain).  The VLT data
  were acquired by means  of the ESO  target of opportunity  programme
  080.D-0008(A).  SK   and DAK acknowledge   support by DFG   grant Kl
  766/16-1.  We are grateful to  Dr.  T.  Szeifert for his suggestions
  on the polarimetric data  reduction.  We thank the anonymous referee
  for constructive comments.
\end{acknowledgements}

\end{document}